# Quantized Friction across Ionic Liquid Thin Films

Alexander M. Smith[1], Kevin R. J. Lovelock[2], Nitya Nand Gosvami[3], Tom Welton[2], and Susan Perkin[1]*

**Affiliations:**

[1] Department of Chemistry, Physical and Theoretical Chemistry Laboratory, University of Oxford, South Parks Road, Oxford OX1 3QZ, U.K.

[2] Department of Chemistry, Imperial College London, London SW7 2AZ, UK

[3] Department of Mechanical Engineering and Applied Mechanics, University of Pennsylvania, 220 S. 33rd St., Philadelphia, PA 19104, USA.

*Correspondence to: susan.perkin@chem.ox.ac.uk

**Ionic liquids – salts in the liquid state under ambient conditions – are of great interest as precision lubricants. Ionic liquids form layered structures at surfaces, yet it is not clear how this nano-structure relates to their lubrication properties. We measured the friction force between atomically smooth solid surfaces across ionic liquid films of controlled thickness in terms of the number of ion layers. Multiple friction-load regimes emerge, each corresponding to a different number of ion layers in the film. In contrast to molecular liquids, the friction coefficients differ for each layer due to their varying composition.**

The importance of friction, and its reduction by lubrication of the sliding surfaces, has been recognized since ancient times and yet some of the most fundamental questions in this area remain unresolved[1, 2]. The classical laws of friction and lubrication do not apply at the nano-scale, where molecular interactions and arrangements determine the surface forces[2-4]. The current renaissance in the study of friction mechanisms is motivated by the need to understand and control shear motion and lubrication in microscopic devices and between nano-scale objects[5-9]. Of particular significance in microscopic systems, or when the external load is relatively small, is the adhesion contribution to the measured friction force[9-12]. The presence of a thin liquid film between two solid surfaces can dramatically modify the friction and adhesion forces between them, and a particularly interesting case arises when the liquid molecules are arranged in layers between the surfaces leading to multiple adhesion values corresponding to different numbers of ion layers in the film[13-15].

Ionic liquids[16] are a particularly attractive class of fluids for precision tribological applications due to their low melting point and wide liquid range, extremely low volatility, high thermal stability, lubrication and wear-protection characteristics, and electrical and thermal conductivity[17-19]. The interfacial structure of ionic liquids at charged surfaces consists of alternating positive and negative ion layers extending several layers into the bulk[20, 21]. This alternating cation/anion structure is even more striking when the liquid is confined to a film between two surfaces: measurements of the surface interaction force as a function of surface separation reveal large oscillations, substantially stronger than van der Waals or mean-field electrostatic interactions, corresponding to sequential squeeze-out of layers from between the surfaces. In contrast to molecular liquids, ionic liquids squeeze out two layers at a time – one cation layer and one anion layer – in order to maintain electroneutrality in the film[15, 21-25]. Despite the burgeoning interest in ionic liquids and their intricate nano-structure, very little is currently known about their mechanism of lubrication[26, 27]. To address this, we performed measurements of friction across films of ionic liquid with key features distinct from earlier studies[15, 26, 27]: (i) friction is measured across ionic liquid films of precisely controlled thickness (number of ion layers) and at both positive and negative applied loads; and (ii) friction, adhesion, and real contact area are all known concurrently for each number of ion layers in the film allowing quantitative analysis. The results reveal a separate friction-load relationship for each number of ion layers. This revelation challenges the classical axiom of tribology that there exists a single-valued friction-load law for any particular lubricated system. The result points at new ways to control friction in microscopic systems in general, and to develop ionic liquid lubricants for specialized applications in particular.

The ionic liquid, 1-butyl-1-methylpyrrolidinium *bis*[(trifluoromethane)sulfonyl]imide or $[C_4C_1Pyrr][NTf_2]$ (Fig. 1(a)), was confined between two mica sheets in a Surface Force Balance (SFB) (Fig. 1(b), [28]). The mica sheets are each ~ 1 cm$^2$ area and ~ 2 μm thickness, atomically smooth and step-free over the whole area, mounted in crossed-cylinder orientation with radius of curvature $R \sim 0.01$ m. The thickness of the liquid film between the sheets was measured with sub-molecular resolution using optical interference according to a method described in detail earlier[29]. We measured the normal interaction force, $F_N$, as a function of mica-mica surface separation, $D$, and the lateral force, $F_L$, between them (as a function of $D$ and $F_N$) which arises when one surface is translated by a lateral displacement $x_S$ at velocity $v_S$. The measured friction responses contained certain features which are defined here as follows (and indicated in Fig. 2): If the film supports a finite stress then during initial translation $F_L$ will increase up to a *yield* force, $F_{S,y}$. If shearing then proceeds by smooth sliding, the lateral force required to maintain sliding is the *kinetic friction* force, $F_{S,k}$. If shearing proceeds by stick-slip, $F_L$ ranges between a minimum of $F_{S,k}$ at the stick-point and a maximum of the *static friction* force, $F_{S,st}$, at the slip-point. If $F_{S,y} > F_{S,k}$ (or $F_{S,y} > F_{S,st}$ in the case of stick-slip) then a *stiction spike* is observed, of height $F_{S,y} - F_{S,k}$.

The profile of $F_N$ with $D$ is an oscillatory function (Fig. 1(c)) due to the packing of the ions into layers and the sequential squeeze-out of pairs of ion layers, as expected for ionic liquids: in the case of $[C_4C_1Pyrr][NTf_2]$ between negatively charged surfaces the stable film structures contain odd-integer numbers of ion layers[24], which we label as $i = 3, 5, 7,$ or $9$, as shown schematically in the insets to Fig.1(d). As a result of the oscillatory force-law between the surfaces across the liquid it is possible to vary $F_N$ – over a certain range including positive and negative values – while the number of ion layers, $i$, in the film remains constant. By performing this compression/de-compression at constant $i$, and at the same time applying lateral displacement to the top surface and detecting the resulting $F_L$, we were able to measure $F_{S,k}$ as a function of $F_N$ for each number of ion layers between the surfaces. The result reveals quantized friction regimes (Fig. 1(d)). Thus there exists a friction-load relationship for each value of $i$, and the friction is multi-valued for each single value of load. This latter point is illustrated in Fig. 1(e) where closely similar values of $F_N$ give rise to different values of $F_{S,y}$ and $F_{S,k}$ depending on the number of ion layers.

Each of the quantized friction regimes follows an approximately linear $F_{S,k}$ vs. $F_N$ relationship with non-zero $F_{S,k}$ measured at zero (and negative) values of $F_N$. The friction at zero applied load originates in the contribution from the adhesion force between the surfaces, $F_{adh}$, which is different for each $i$, as seen from the $F_N$ minima in Fig. 1c. $F_{S,k}$ can be expressed as a linear combination of load-dependent and adhesion-dependent contributions[10]:

$$F_{S,k}^i = \mu^i F_N + \alpha^i \left(\frac{A^i \cdot F_{adh}^i}{\pi d^i R}\right) \qquad \text{Eq. 1}$$

where $\mu$ is the load-controlled friction coefficient, $\alpha$ is the adhesion-controlled friction coefficient, $A^i$ is the flat contact area, $d^i$ is the distance between the shearing liquid layers, and superscripts $i$ denote values for a film containing $i$ layers. In our experiments we directly measured $F_{adh}^i$, $F_{S,k}^i$, $R$, and $d^i$ whilst controlling $F_N$. $A^i$ is related to $F_N$ and $F_{S,k}^i$ by the JKR equation and was determined after $in$-$situ$ calibration of the elastic modulus[14]. Thus it was possible to determine values of the coefficients $\mu^i$ and $\alpha^i$ for each discrete film thickness. We found that for the films consisting of 9, 7 and 5 ion layers $\mu^i$ increases gradually in the range 0.082 – 0.172 (±0.008), whereas for $i = 3$ the gradient is much steeper giving $\mu^{i=3} = 0.812 \pm 0.090$. The adhesion contribution to the friction gives coefficients with closely similar values for all $i$; $\alpha^i = 0.19 \pm 0.04$ [30]. The increase in $\mu^i$ as $i$ decreases is likely due to the greater degree of interlocking of the layers for small $i$: the layers closer to the mica surfaces are more ordered, and have greater anion/cation excess concentration, and therefore have greater inter-layer attractions for the same value of contact area and a greater activation barrier for 'unlocking' the surfaces to allow shear.

To obtain further insight into the molecular reorganization taking place during shear we recorded high resolution traces of $F_L$ as a function of lateral displacement of the top surface for a range of $F_N$ and $v_S$. Figures 2 and 3 show examples for $i = 7$; qualitatively similar results were observed for $i =$ 9, 7, 5, and 3. Over the range of $F_N$ and $v_S$ studied the films exhibit a well defined yield point, with $F_{S,y}$ increasing in magnitude as $F_N$ increases, and stiction spikes at the onset of sliding (Fig. 2). At low $v_S$, shearing of the film then progresses by way of a series of stick-slip cycles characterized by saw-tooth traces (Fig. 3(a)). At a critical $v_S$ (Fig. 3(b)) the stick-slip disappears and above this velocity the film shears with a smooth sliding motion (Fig. 3(c)). It was notable that $F_{S,k}$ is independent of $v_S$, whilst $F_{S,y}$ is weakly dependent on $v_S$. Once the direction of applied lateral motion is reversed, similar behavior is observed in the opposite direction.

The clear yield points indicate solid-like behavior of the ionic liquid confined to films of $i \leq 9$ ion layers. Subsequent stick-slip behavior observed at lower $v_S$ is most often attributed to a series of freeze-melt transitions[5, 31]. However in the case of ionic liquids a full-film melt is unlikely due to the high Coulombic energy barrier of ions swapping between layers, so the stick-slip is likely to involve either interlayer slips[32] or intra-layer (two dimensional) melting. Furthermore, the particularly high values of $F_{S,k}^{i=3}$ compared to $F_{S,k}^{i>3}$ indicates that mid-film cation layers – the structural feature present in films of greater thickness but not in the $i = 3$ film – may shear-melt or slip at lower stress than the other layers and thus be responsible for the lower yield force of the thicker films.

The *quantized friction* presented here is a direct result of the multiple discrete values of the adhesion-controlled contribution to the total friction, and is expected to be general for liquids which form layered structures in thin films. Indeed, experiments in our laboratory with a range of other ionic liquids have revealed quantized regimes corresponding to the layer structure in every case. The result rationalizes previous reports of discontinuities in the friction across molecular liquids when the number of layers changes[10, 14], here demonstrating multiple quantized friction regimes with varying friction coefficients and quantifying the effects of adhesion on the overall friction force. This result will be pivotal in explaining the boundary lubrication of rough surfaces where the liquid film is of varying thicknesses across the contact zone: the total friction is made up of differently weighted contributions from each film thickness despite uniform applied load. Thus the quantized friction regimes resolved here using atomically-smooth surfaces will provide a link between single-asperity and rough-contact friction.

**Acknowledgments:** This work was supported by The Leverhulme Trust (F/07 134/DK and F/07 134/DN), Taiho Kogyo Tribology Research Foundation, Weizmann UK, Infineum UK, The Office of Naval Research (N00014-10-1-0096) and the EPSRC (EP/J015202/1). Andrew Dolan is thanked for synthesis of the ionic liquid used in this work.

**Supplemental Material**

**Materials:** $[C_4C_1Pyrr][NTf_2]$ was synthesized according to existing literature methods[33] at Imperial College London. The liquid was not treated by column chromatography, since this has been shown to introduce particulate impurities[34]. Ionic liquids were thoroughly washed with water then dried *in vacuo* ($10^{-2}$ mbar, 70 °C) for 24 hours prior to injection into the SFB apparatus. Karl Fischer titration showed the water levels were <50 ppm prior to injection. SFB friction measurements were carried out using ~100 μL of ionic liquid suspended between two freshly cleaved and atomically smooth (step-free) muscovite mica sheets and at T = 22±1°C. The oscillatory normal forces provide strong evidence for the dry and clean nature of the ionic liquid; these are seen to disappear upon introducing humidity to the environment. The data in Fig. 1**d** was measured over 5 different experiments (different pairs of mica sheets and different ionic liquid samples), and at multiple contact spots on the mica for each experiment.

**Determination of the load and adhesion friction coefficients:** Linear fits to the measured $F_{S,k}$ and $F_N$ data (Fig. 1d) for each $i$ were used to obtain $\mu^i$ (gradient) and $F_{S,k}$ at $F_N = 0$ (intercept). $d^i$, the distance between an anion layer and the adjacent cation layer, as obtained from the half the distance between adjacent minima in the oscillatory force law (Fig. 1c), and $F_{adh}$ was measured from the force required to pull apart the surfaces (minima of the oscillations in Fig. 1c). $R$ was measured from the optical interferometry. $K$ was calibrated by applying much larger normal forces and the corresponding large flattened area measured in the interference fringes was fitted using the JKR equation. The values used to then calculate $\alpha^i$ using Eq. 1 (and the associated propagated errors) are shown below.

| $i$ | $\mu^i$ | error | $F_{adh}$ (μN) | error | $F_S$ at $F_N=0$ | error | $d^i$ (nm) | error | $A_0$ (μm$^2$) | $\alpha^i$ | error |
|---|---|---|---|---|---|---|---|---|---|---|---|
| 9 | **0.082** | 0.008 | 12.28 | 3.07 | 1.22 | 0.37 | 0.50 | 0.05 | 12.76 | **0.18** | 0.13 |
| 7 | **0.125** | 0.006 | 25.32 | 6.33 | 7.96 | 0.57 | 0.45 | 0.05 | 20.67 | **0.22** | 0.09 |
| 5 | **0.172** | 0.008 | 52.71 | 13.18 | 28.67 | 1.35 | 0.45 | 0.05 | 33.71 | **0.23** | 0.08 |
| 3 | **0.812** | 0.090 | 99.67 | 24.92 | 49.79 | 2.59 | 0.50 | 0.05 | 51.54 | **0.15** | 0.09 |

**References and Notes:**

[1]  B. N. J. Persson, *Sliding Friction* (Springer-Verlag, 1998).

[2]  M. Urbakh *et al.*, Nature **430** (2004).

[3]  B. Bhushan, J. Israelachvili, and U. Landman, Nature **374** (1995).

[4]  Y. Mo, K. T. Turner, and I. Szlufarska, Nature **457** (2009).

[5]  P. A. Thompson, and M. O. Robbins, Science **250** (1990).

[6]  J. Y. Park *et al.*, Science **313** (2006).

[7]  A. Socoliuc *et al.*, Science **313** (2006).

[8]  C. Lee *et al.*, Science **328** (2010).

[9]  R. Guerra *et al.*, Nat Mater **9** (2010).

[10]  J. Israelachvili, Y. L. Chen, and H. Yoshizawa, Journal of Adhesion Science and Technology **8** (1994).

[11]  Z. Deng *et al.*, Nat Mater **11** (2012).

[12]  B. Luan, and M. O. Robbins, Nature **435** (2005).

[13]  R. Horn, and J. Israelachvili, Journal of Chemical Physics **75** (1981).

[14]  E. Kumacheva, and J. Klein, Journal of Chemical Physics **108** (1998).

[15]  S. Perkin, T. Albrecht, and J. Klein, Physical Chemistry Chemical Physics **12** (2010).

[16]  J. P. Hallett, and T. Welton, Chem Rev **111** (2011).

[17]  M. Palacio, and B. Bhushan, Tribology Letters **40** (2010).

[18]  F. Zhou, Y. Liang, and W. Liu, Chemical Society Reviews **38** (2009).

[19]  A. Somers *et al.*, Lubricants **1** (2013).

[20]  M. Mezger *et al.*, Science **322** (2008).

[21]  R. Atkin, and G. G. Warr, Journal of Physical Chemistry C **111** (2007).

[22]  K. Ueno *et al.*, Physical Chemistry Chemical Physics **12** (2010).

[23]  I. Bou-Malham, and L. Bureau, Soft Matter **6** (2010).

[24]  A. M. Smith *et al.*, The Journal of Physical Chemistry Letters  (2013).

[25]  S. Perkin *et al.*, Chemical Communications **47** (2011).

[26]  J. Sweeney *et al.*, Physical Review Letters **109** (2012).

[27]  O. Werzer *et al.*, Physical Chemistry Chemical Physics  (2012).

[28]  See Supplemental Material at [URL] for methods

[29]  S. Perkin *et al.*, Langmuir **22** (2006).

[30]  See Supplemental Material at [URL] for all values

[31]  M. Schoen *et al.*, Science **245** (1989).

[32] Y. Lei, and Y. Leng, Physical Review Letters **107** (2011).

[33] M. A. Ab Rani *et al.*, Physical Chemistry Chemical Physics **13** (2011).

[34] F. Endres, S. Z. El Abedin, and N. Borissenko, Z. Phys. Chem. **220** (2006).

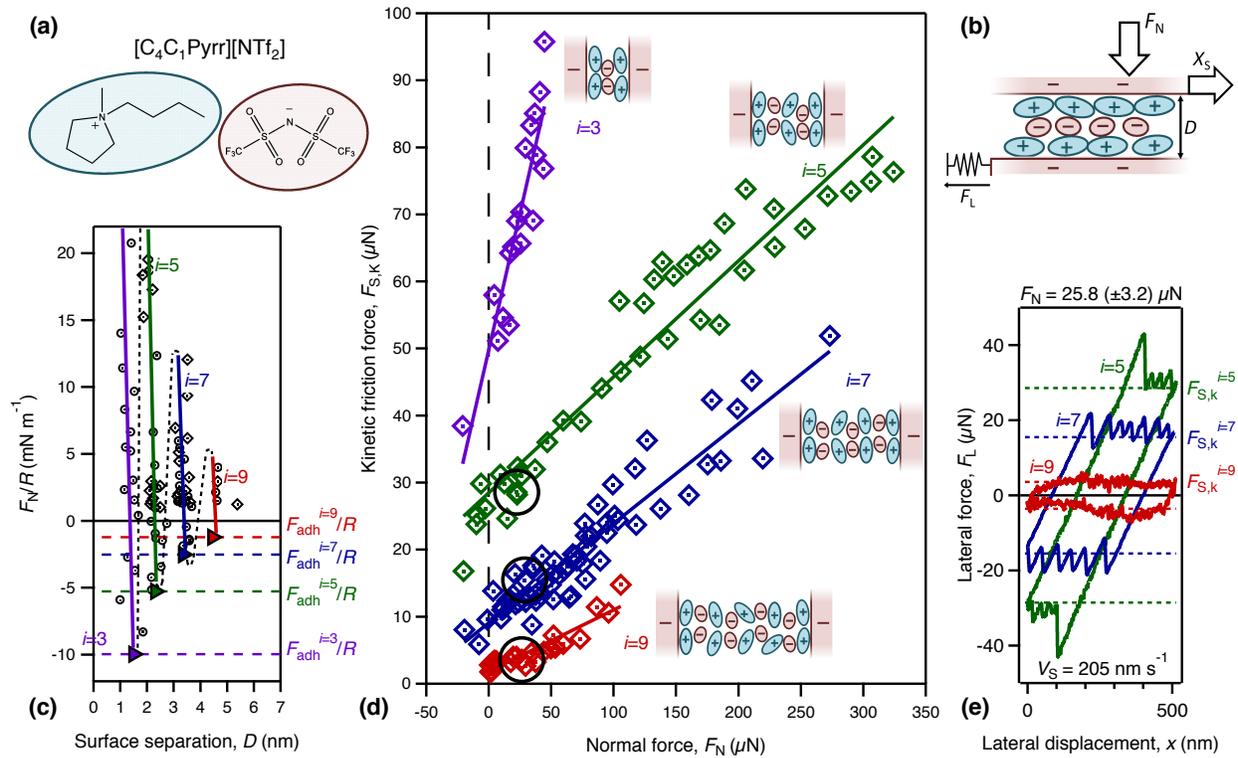

**Figure 1**: **Normal and shear forces between mica sheets across films of [C$_4$C$_1$Pyrr][NTf$_2$].** (**a**) The cation and anion structures for the ionic liquid [C$_4$C$_1$Pyrr][NTf$_2$]. (**b**) Schematic showing the system of ionic liquid arranged in layers between negative mica surfaces. (**c**) $F_N$ vs. $D$ showing the oscillatory force law due to sequential squeeze-out of pairs of ion layers, labeled with the number of ion layers at each distance and the corresponding values of $F_{adh}/R$; data reproduced with permission[24]. (**d**) $F_{S,k}$ vs. $F_N$ measured for different numbers of ion layers, $i$, in the film. The lines are linear fits to the data according to equation (1). Inset diagrams indicate the liquid structure for each regime, and the circled data points are those corresponding to the traces in (**e**). (**e**) Friction loops showing $F_L$ vs. lateral displacement, $x_S$, showing three examples measured at similar $F_N$ and $v_S$ but different numbers of ion layers ($i$ = 9, 7, and 5).

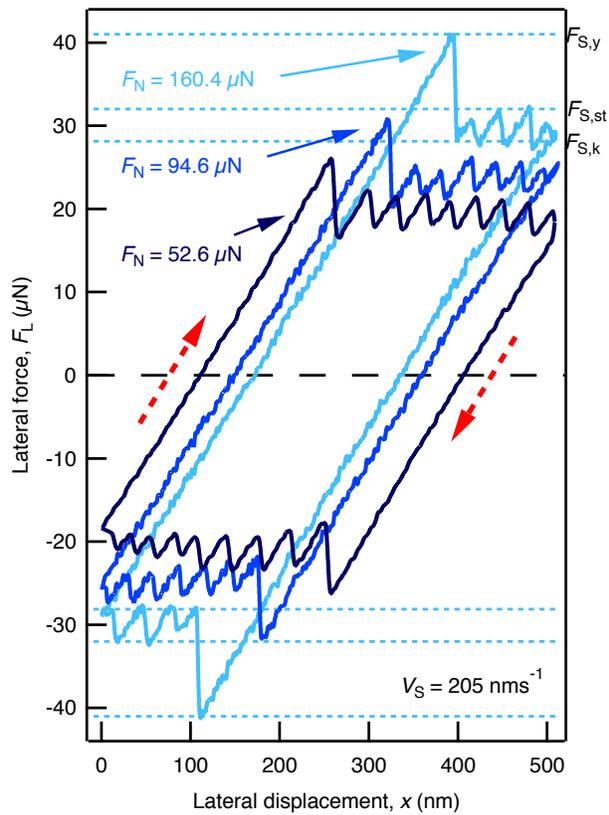

**Figure 2: Friction loops, showing $F_L$ as a function of $x_S$ at a constant film thickness of $i = 7$ and for three different values of $F_N$.** The velocity is constant at $v_S = 205$ nm/s, and the back-and-forth displacement is in the directions indicated by (red) dashed arrows. Friction features $F_{S,y}$, $F_{S,k}$, and $F_{S,st}$ are defined for the $F_N = 160.4$ μN trace.

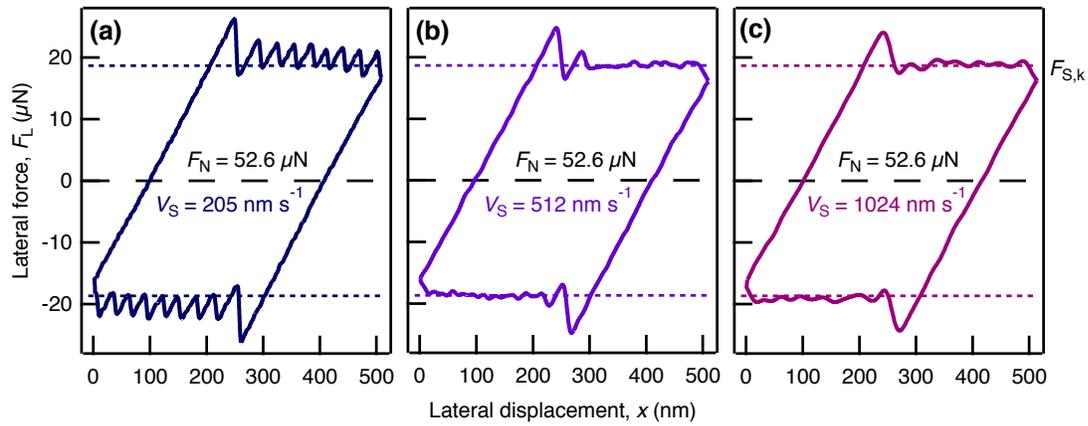

**Figure 3: Friction loops for varying $v_S$ at constant $F_N$.** **(a)** $v_S = 205$ nm s$^{-1}$; **(b)** $v_S = 512$ nm s$^{-1}$; **(c)** $v_S = 1024$ nm s$^{-1}$ ; each measured at $F_N = 52.6$ $\mu$N. $F_{S,y} > F_{S,k}$ leading to a stiction spike in all cases, but only at slower $v_S$ are regular stick-slip cycles observed. At the critical velocity, here 512 nm s$^{-1}$, one stick-slip cycle is followed by smooth sliding. At higher $v_S$ smooth sliding occurs after the initial spike.